\documentclass[11pt,a4paper]{article}
\usepackage{jheppub}

\usepackage{graphicx,epsfig}
\usepackage{amsmath}
\usepackage{amssymb,color}

\newcommand{\be}{\begin{eqnarray}}
\newcommand{\ee}{\end{eqnarray}}
\newcommand{\bea}{\begin{eqnarray}}

\newcommand{\eea}{\end{eqnarray}}

\title{\textbf Renormalized AdS action and Critical Gravity}

\author[a]{Olivera Miskovic,}
\author[b]{Minas Tsoukalas}
\author[c]{and Rodrigo Olea}

\affiliation[a]{Instituto de F\'{\i}sica, Pontificia Universidad Cat\'{o}lica de Valpara\'\i so,\\
Casilla 4059, Valpara\'{\i}so, Chile}
\affiliation[b]{Departamento de Ciencias F\' isicas,
Universidad Andres Bello, \\ Rep\' ublica 220, Santiago, Chile}
\affiliation[c]{Centro de Estudios Cient\' ificos (CECs),\\
Casilla 1469, Valdivia, Chile}
\date{\today }

\emailAdd{olivera.miskovic@ucv.cl}
\emailAdd{minasts@cecs.cl}
\emailAdd{rodrigo.olea@unab.cl}

\abstract{ It is shown that the renormalized action for AdS gravity in even spacetime dimensions is equivalent
 --on shell-- to a polynomial of the Weyl tensor, whose first non-vanishing term is proportional to $Weyl^2$.
Remarkably enough, the coupling of this last term coincides with the one that appears in Critical Gravity.}


\begin{document}
\maketitle
%\flushbottom

%%%%%%%%%%%%%%%%%%%%%%%%%%%%%%%%%%%%%%%%%%%%%%%%%%%%%%%%%%%%%%%%%%%%%%%%%%%
\section{Introduction}
%%%%%%%%%%%%%%%%%%%%%%%%%%%%%%%%%%%%%%%%%%%%%%%%%%%%%%%%%%%%%%%%%%%%%%%%%%%

Despite the huge success of General Relativity, it is still an insufficient
theory since it does not incorporate quantum phenomena. The most promising
candidate up to date to achieve such a goal seems to be String Theory
that predicts non-linear terms in the curvature in the low-energy limit.
 In four dimensions, if one wants to modify the
Einstein-Hilbert action such that Ostrogradski ghosts  are absent \cite{ostro}
while maintaining a massless spin 2 degree of freedom, the only possibility
is adding a cosmological constant.
Including higher-order curvature terms is another
possible modification.
These terms can be added to the action in a
high-energy regime, rendering the theory perturbatively
renormalizable in absence of the cosmological constant
\cite{Stelle:1976gc,Stelle:1977ry}, with the drawback of appearance
of ghosts in the form of massive spin-2 modes.

The issue of higher-order terms in the four-dimensional action has
been recently revisited from the point of view of Critical Gravity
\cite{Lu:2011zk}, where quadratic terms in the Ricci tensor and the Ricci scalar were
considered on top of the Einstein-Hilbert action with negative cosmological
constant. In that line of reasoning, the presence of the
cosmological constant is crucial. Indeed, perturbative analysis around
anti-de Sitter (AdS) vacuum leads to constraints on the parameters of the theory when
the massive spin-2 mode is rendered massless.
The coupling constants are further restricted by the cancelation of the scalar excitation
and they get tuned  with the inverse of the cosmological constant. The
on-shell energy of the remaining massless spin-2 mode becomes
zero and so do the mass and the entropy of the black holes of the theory.

However, the presence of a double pole structure in the theory allows for
logarithmic modes, which ruins unitarity of the theory
\cite{Bergshoeff:2011ri,Alishahiha:2011yb,Gullu:2011sj,AyonBeato:2009yq,Porrati:2011ku,Lu:2011ks}.
Proper boundary conditions can eliminate these logarithmic
modes, because their fall-off is slower than the one of modes.
As a result, obtained four-dimensional unitary gravitational
action contains only one term, that is the square of the Weyl tensor, and
also the Gauss-Bonnet term which, in four dimensions, does not contribute
to the field equations \cite{Lovelock:1971yv}.

The concept of Critical Gravity with quadratic terms in the
curvature was generalized to higher dimensions. Apart from Ricci-squared
and Ricci scalar-squared contributions, the Gauss-Bonnet term becomes dynamical
in the higher-dimensional setup  \cite{Lovelock:1971yv,Boulware:1985wk}.
In the case that there is a single vacuum, a suitable choice of the couplings eliminates
both the scalar mode and the mass of the massive mode. This particular point in
the space of parameters provides a reconciling picture to deal with the problem that,
in general, the mass of the spin-2 excitation and the one of the black holes have
opposite signs: both have vanishing on-shell energy in Critical Gravity. As a result,
the theory can be written in terms of the Einstein-Hilbert action and the square of
the Weyl tensor \cite{Deser:2011xc},
\begin{equation}
I_{\text{CG}}=\frac{1}{16\pi G}\int d^D x\sqrt{-g}\,\left( R-2\Lambda
-\gamma_{_\text{CG}} \,W^{\mu \nu \alpha \beta }W_{\mu \nu \alpha \beta }\right) \,,
\label{CGA}
\end{equation}
where the cosmological constant expressed in terms of AdS radius is $\Lambda =-(D-1)(D-2)/2\ell^2$.
Here, the Weyl-squared term reads
\begin{equation}
W^{\mu \nu \alpha \beta }W_{\mu \nu \alpha \beta }=R^{\mu \nu \alpha \beta
}R_{\mu \nu \alpha \beta }-\frac{4}{D-2}\,R^{\mu \nu }R_{\mu \nu }
+\frac{2}{(D-1)(D-2)}\,R^2\,,
\end{equation}
with the  coupling constant for Critical Gravity given by
\begin{equation}
\gamma_{_\text{CG}} =-\frac{(D-1)(D-2)}{8\Lambda (D-3)}\,.
\label{gammaCG}
\end{equation}
Note that the flat limit ($\Lambda =0$) is not well-defined.

Other type of Critical Gravity with quadratic curvature contributions was discussed in Ref.\cite{Kan:2013moa}, but it does not
include the Riemann square term.
Extensions of Critical Gravity with cubic-curvature invariants
were studied in Ref.\cite{Sisman:2011gz}.

In a different line of development, it was shown that the AdS action in four dimensions
evaluated on-shell is \cite{Mis-Ole,Maldacena:2011mk}
\begin{equation}
I_{\text{ren}}=\frac{\ell^2}{64\pi G}\int d^4 x\sqrt{-g} \,W^{\mu \nu \alpha \beta }W_{\mu \nu \alpha \beta }\,,
\label{Iren4D}
\end{equation}
where $I_{\text{ren}}$ is the action properly renormalized by the addition of counterterms \cite{de Haro:2000xn}.
The proof in Ref.\cite{Mis-Ole} makes use of the renormalizing effect of topological invariants, as the addition
of Gauss-Bonnet is equivalent to Holographic Renormalization procedure in asymptotically
AdS gravity. Then, the bulk action becomes manifestly the one of Conformal Gravity \cite{Adler:1982ri,Mannheim:2011ds,tHooft:2011aa,boul}
for Einstein spacetimes.

From a different point of view, the use of appropriate boundary conditions
in the infrared regime of the theory led to the same conclusion in Ref.\cite{Maldacena:2011mk}.
Curiously enough, the relation between the cosmological constant and the coupling of Weyl square
is exactly the same one that appears in Critical Gravity, while the boundary conditions
are those which eliminate the logarithmic modes \cite{Lu:2011ks}.

The remarkable feature of Eq.(\ref{Iren4D}) is that the coefficient of $Weyl^2$ term
is exactly the same as the coupling $\gamma_{_\text{CG}}$ that appears in the Critical Gravity action (\ref{CGA}).

In this paper, we extend this result to higher even dimensions along the line of the argument presented in
Ref.\cite{Mis-Ole}. Indeed, we show that the renormalized AdS action becomes on-shell a polynomial of the Weyl
tensor, whose first term is always $\gamma_{_\text{CG}}Weyl^2$.

%%%%%%%%%%%%%%%%%%%%%%%%%%%%%%%%%%%%%%%%%%%%%%%%%%%%%%%%%%%%%%%%%%%%%%%%%%%
\section{Renormalized AdS action in even dimensions, Kounterterms and
topological invariants}
%%%%%%%%%%%%%%%%%%%%%%%%%%%%%%%%%%%%%%%%%%%%%%%%%%%%%%%%%%%%%%%%%%%%%%%%%%%

In the context of AdS/CFT correspondence, the gravity action requires the
addition of a counterterm series $\mathcal{L}$, which are surface terms
constructed only with intrinsic quantities of the boundary, in order to
cancel the divergences that appear in the asymptotic region. In doing so,
the action and its variation are functionals only of a given conformal
structure at the boundary, $[g_{(0)ij}]$, which is the source of the dual CFT.

Throughout the paper, we will use the radial foliation of the manifold $M$
\begin{equation}
ds^{2}=g_{\mu\nu}\,dx^{\mu}dx^{\nu}=N^{2}\left(  \rho\right)
\,d\rho^{2}+h_{ij}(\rho,x)\,dx^{i}dx^{j}\,,\label{radial foliation}
\end{equation}
where $x^{i}$  and $h_{ij}$ are the coordinates and the metric at the boundary $\partial M$,
respectively.

In the even-dimensional case, the renormalized AdS action reads
\cite{de Haro:2000xn}
\begin{eqnarray}
I_{\text{ren}} &=&I_{\text{EH}}-\frac{1}{8\pi G}\int\limits_{\partial M}d^{2n-1}x\,
\sqrt{-h}\,K+\frac{1}{8\pi G}\int\limits_{\partial M}d^{2n-1}x\,\sqrt{-h}\,\left[ \frac{%
2n-2 }{\ell }+\right.   \notag \\
&&\left. +\frac{\ell \mathcal{R}}{2(2n-3)}+\frac{\ell ^{3}}{2(2n-3)^{2}(2n-5)%
}\left( \mathcal{R}_{ij}\mathcal{R}^{ij}-\frac{(2n-1)}{4(2n-2)}\mathcal{R}%
^{2}\right) +\cdots \right] \,,
\end{eqnarray}
where the second term in the first line is the Gibbons-Hawking-York (GHY)
term, which ensures a well-posed action principle for $\delta h_{ij}=0$ as a
boundary condition.

We have defined the extrinsic curvature as
\begin{equation}
\label{eq:extrcurv}
K_{ij}=\frac{1}{2N}\,\partial_{\rho}h_{ij}\,,
\end{equation}
and $\mathcal{R}_{jkl}^{i}(h)$ is the intrinsic curvature.

However, in asymptotically AdS (AAdS) spacetimes, the behavior of $\delta h_{ij}$ at the boundary
is divergent such that --strictly speaking-- the counterterms are also
needed for the variational problem \cite{Papadimitriou:2005ii}.
Variations of the extrinsic curvature are equally ill-defined at the
boundary, because the leading order in the expansion of $\delta K_{ij}$ is
the same as the leading order in $\delta h_{ij}$ due to the conformal
structure of the boundary. This fact motivates the inclusion of counterterms
which depend on the extrinsic curvature instead of the standard series given
above.

Indeed, extrinsic counterterms for Einstein-Hilbert AdS gravity were
proposed in Refs.\cite{Olea:2005gb} and \cite{Olea:2006vd}.

In $D=2n$ dimensions, we consider the AdS action, renormalized with the
addition of this alternative counterterm series (a.k.a. Kounterterms)
\begin{equation}
\tilde{I}_{\text{ren}}=I_{\text{EH}}+c_{2n-1}\int\limits_{\partial
M}d^{2n-1}x\,B_{2n-1}(h,K,\mathcal{R})\,,  \label{Irentilde}
\end{equation}
where the boundary terms in the even-dimensional case are given by
\begin{eqnarray}
B_{2n-1}(h,K,\mathcal{R}) &=&2n\,\sqrt{-h}\int\limits_{0}^{1}dt\,\delta _{%
\left[ j_{1}\cdots j_{2n-1}\right] }^{\left[ i_{1}\cdots i_{2n-1}\right]
}\,K_{i_{1}}^{j_{1}}\left( {\frac{1}{2}\,}\mathcal{R}%
_{i_{2}i_{3}}^{j_{2}j_{3}}(h)-t^{2}K_{i_{2}}^{j_{2}}K_{i_{3}}^{j_{3}}\right)
\times \cdots  \notag \\
&&\qquad \qquad \cdots \times \left( {\frac{1}{2}\,}\mathcal{R}%
_{i_{2n-2}i_{2n-1}}^{j_{2n-2}j_{2n-1}}(h)-t^{2}K_{i_{2n-2}}^{j_{2n-2}}K_{i_{2n-1}}^{j_{2n-1}}\right) \,,
\label{B_2n}
\end{eqnarray}
with a coefficient given in terms of the AdS radius as
\begin{equation}
c_{2n-1}=\frac{1}{16\pi G}\,\frac{\left( -1\right) ^{n}\,\ell ^{2n-2}}{%
n\left( 2n-2\right) !}\,.
\end{equation}
Above, the totally antisymmetric Kronecker delta of order $m$ is defined as the determinant
of single-index Kronecker deltas.

A relation between this extrinsic counterterm series and the standard one
was sketched in Ref.\cite{Mis-Ole}. Here, a comparison to standard
renormalization procedure is given in more detail.

We start by adding and subtracting the GHY term,
\begin{equation}
\tilde{I}_{\text{ren}}=I_{\text{EH}}-\frac{1}{8\pi G}\int\limits_{\partial M}d^{2n-1}x%
\sqrt{-h}\,K+\int\limits_{\partial M}d^{2n-1}x\,\mathcal{L}(h,K,\mathcal{R})\,,
\end{equation}
where
\begin{equation}
\mathcal{L}(h,K,\mathcal{R})=c_{2n-1}B_{n-1}-GHY.
\end{equation}
It is easier to manipulate $\mathcal{L}(h,K,\mathcal{R})$ if we write down
the last term in the above relation as a totally antisymmetric object, that
is,
\begin{eqnarray}
\mathcal{L}(h,K,\mathcal{R}) &=&\frac{\left( -1\right) ^{n}\ell ^{2n-2}
\sqrt{-h}}{8\pi G(2n-2)!}\,\delta _{\left[ j_{1}\cdots j_{2n-1}\right] }^{\left[
i_{1}\cdots i_{2n-1}\right] }\,K_{i_{1}}^{j_{1}}\int\limits_{0}^{1}dt\ \left[
\left( \frac{1}{2}\mathcal{R}%
_{i_{2}i_{3}}^{j_{2}j_{3}}-t^{2}K_{i_{2}}^{j_{2}}K_{i_{3}}^{j_{3}}\right)
\times \cdots \right.  \notag \\
&&\quad \cdots \left. \times \left( \frac{1}{2}
\mathcal{R}_{i_{2n-2}i_{2n-1}}^{j_{2n-2}j_{2n-1}}-t^{2}K_{i_{2n-2}}^{j_{2n-2}}K_{i_{2n-1}}^{j_{2n-1}}\right) +
\frac{(-1)^{n}}{\ell ^{2n-2}}\,\delta _{i_{2}}^{j_{2}}\cdots \delta
_{i_{2n-1}}^{j_{2n-1}}\right] .  \label{LhKR}
\end{eqnarray}
On the other hand, for any AAdS spacetime, the asymptotic
expansion of the extrinsic curvature is given by
\begin{equation}
K_{j}^{i}=\frac{1}{\ell }\,\delta _{j}^{i}+\ell S_{j}^{i}(h)+\mathcal{O}(%
\mathcal{R}^{2})\,,  \label{Kexpansion}
\end{equation}
up to second-derivative terms. Here, the quantity $S_{j}^{i}$ is the
Schouten tensor of the boundary metric, that is,
\begin{equation}
S_{j}^{i}(h)=\frac{1}{2n-3}\left( \mathcal{R}_{j}^{i}(h)-\frac{1}{4(n-1)}\,
\delta _{j}^{i}\mathcal{R}(h)\right) .
\end{equation}
This expansion implies that $\mathcal{L}(h,K,\mathcal{R})$ is expressible in
terms of intrinsic quantities of the boundary at least up to quadratic
terms in the curvature. Direct substitution of relation (\ref{Kexpansion})
in the general boundary term (\ref{LhKR}) produces a rather complicated
expression
\begin{eqnarray}
\mathcal{L}(h,K,\mathcal{R}) &=&\frac{(-1)^n\ell^{2n-2}}{8\pi G(2n-2)!}\,
\sqrt{-h}\,\delta _{\left[ j_{1}\cdots j_{2n-1}\right] }^{\left[i_{1}\cdots i_{2n-1}\right]}
\left( \frac{1}{\ell }\,\delta_{i_1}^{j_1}+\ell S_{i_1}^{j_1}+\cdots\right) \times  \notag \\
&&\times \int\limits_{0}^{1}dt\ \left[ \left( -\frac{t^2}{\ell^2}\,\delta
_{i_{2}}^{j_{2}}\delta _{i_{3}}^{j_{3}}+\frac{1}{2}\left(
\mathcal{R}_{i_{2}i_{3}}^{j_{2}j_{3}}-4t^{2}S_{i_{2}}^{j_{2}}
\delta_{i_{3}}^{j_{3}}\right) -t^{2}\ell
^{2}S_{i_{2}}^{j_{2}}S_{i_{3}}^{j_{3}}+\cdots\right) \times \cdots \right. \notag \\
&&\left.  \times \left( -\frac{t^2}{\ell^2}\,\delta_{i_{2n-2}}^{j_{2n-2}}
\delta _{i_{2n-1}}^{j_{2n-1}}+\frac{1}{2}\left(\mathcal{R}_{i_{2n-2}i_{2n-1}}^{j_{2n-2}j_{2n-1}}-4t^{2}
S_{i_{2n-2}}^{j_{2n-2}}\delta_{i_{2n-1}}^{j_{2n-1}}\right) -t^{2}\ell
^{2}S_{i_{2n-2}}^{j_{2n-2}}S_{i_{2n-1}}^{j_{2n-1}}+\cdots\right) \right.  \notag\\
&&+\left.\frac{(-1)^{n}}{\ell ^{2n-2}}\,\delta _{i_{1}}^{j_{1}}\cdots \delta
_{i_{2n-1}}^{j_{2n-1}}\right] .  \label{LhKRexpand}
\end{eqnarray}
Using the definition of the Weyl tensor of the boundary metric in terms of
the Riemann and the Schouten tensor, and the skew symmetry of its indices,
whenever the boundary Weyl tensor enters in a totally antisymmetric formula
as the one above, we have that
\begin{equation}
\delta _{\left[ \cdots j_{p}j_{p+1}\cdots \right] }^{\left[ \cdots
i_{p}i_{p+1}\cdots \right] }\,\mathcal{W}_{i_{p}i_{p+1}}^{j_{p}j_{p+1}}=\delta
_{\left[ \cdots j_{p}j_{p+1}\cdots \right] }^{\left[ \cdots
i_{p}i_{p+1}\cdots \right] }\left( \mathcal{R}%
_{i_{p}i_{p+1}}^{j_{p}j_{p+1}}-4S_{i_{p}}^{j_{p}}\delta
_{i_{p+1}}^{j_{p+1}}\right) .  \label{WRS}
\end{equation}

We use the last relation to eliminate the dependence of the Riemann tensor
in Eq.(\ref{LhKRexpand}), such that it can be rewritten as
\begin{eqnarray}
\mathcal{L} &=&\frac{\left( -1\right) ^{n}\ell ^{2n-2}}{8\pi G(2n-2)!}\,
\sqrt{-h}\,\delta _{\left[ j_{1}\cdots j_{2n-1}\right] }^{\left[ i_{1}\cdots i_{2n-1}
\right] }\left( \frac{1}{\ell }\delta _{i_{1}}^{j_{1}}+\ell
S_{i_{1}}^{j_{1}}+...\right) \times  \notag \\
&&\times \int\limits_{0}^{1}dt\ \left[ \left( -\frac{t^{2}}{\ell ^{2}}\,\delta
_{i_{2}}^{j_{2}}\delta _{i_{3}}^{j_{3}}+\frac{1}{2}
\left( \mathcal{W}_{i_{2}i_{3}}^{j_{2}j_{3}}+4\left( 1-t^{2}\right) S_{i_{2}}^{j_{2}}\delta
_{i_{3}}^{j_{3}}\right) -t^{2}\ell
^{2}S_{i_{2}}^{j_{2}}S_{i_{3}}^{j_{3}}+\cdots\right) \times \cdots \right.
\notag \\
&&\left. \times \left( -\frac{t^{2}}{\ell ^{2}}\,\delta
_{i_{2n-2}}^{j_{2n-2}}\delta _{i_{2n-1}}^{j_{2n-1}}+\frac{1}{2}\left(
\mathcal{W}_{i_{2n-2}i_{2n-1}}^{j_{2n-2}j_{2n-1}}+4\left( 1-t^{2}\right)
S_{i_{2n-2}}^{j_{2n-2}}\delta _{i_{2n-1}}^{j_{2n-1}}\right) -t^{2}\ell
^{2}S_{i_{2n-2}}^{j_{2n-2}}S_{i_{2n-1}}^{j_{2n-1}}+\cdots\right)\right.
\notag \\
&&\left. +\frac{(-1)^{n}}{\ell ^{2n-2}}\,\delta _{i_{2}}^{j_{2}}\cdots \delta
_{i_{2n-1}}^{j_{2n-1}}\right] .
\end{eqnarray}
Notice that the term $\left( -\frac{t^{2}}{\ell ^{2}}\,\delta \delta
+\frac{1}{2}\left( \mathcal{W}+4\left( 1-t^{2}\right) S\delta \right) -t^2\ell
^{2}SS+\cdots\right) $ appears $(n-1)$ times.

The key point to generate the standard counterterm series from the above
formula is to identify the contributions coming from $\mathcal{L}$ as an
expansion in powers of the boundary curvature. Symbolically, the
lowest-order terms in the expansion of the trinomial to the $(n-1)$-th power
are
\begin{eqnarray}
&&\left( -\frac{t^{2}}{\ell ^{2}}\delta \delta +\frac{1}{2}\left( \mathcal{W}%
+4\left( 1-t^{2}\right) S\delta \right) -t^{2}\ell ^{2}SS+\cdots\right) ^{n-1}
\notag \\
&=&\left( -\frac{t^{2}}{\ell ^{2}}\right) ^{n-1}\left( \delta \right)
^{2n-2}+\frac{(n-1)}{2}\left( -\frac{t^{2}}{\ell ^{2}}\right) ^{n-2}\left(
\delta \right) ^{2n-4}\left( \mathcal{W}+4\left( 1-t^{2}\right) S\delta
\right)  \notag \\
&&+\frac{(n-1)(n-2)}{8}\left( -\frac{t^{2}}{\ell ^{2}}\right) ^{n-3}\left(
\delta \right) ^{2n-6}\left( \mathcal{W}+4\left( 1-t^{2}\right) S\delta
\right) ^{2} \notag \\
&&-(n-1)\left( -\frac{t^{2}}{\ell ^{2}}\right) ^{n-2}t^{2}\left( \delta
\right) ^{2n-4}SS+\cdots\,.
\end{eqnarray}
The term with no curvatures comes just from the multiplication of Kronecker
deltas,\footnote{
If $N$ is the range of indices, a contraction of $k$ indices in the
Kronecker delta of order $m$ produces a delta of order $m-k$,
$$
\delta _{[ j_{1}\cdots j_{k}\cdots j_{m}] }^{[i_{1}\cdots i_{k}\cdots i_{m}]}\,
\delta _{i_{1}}^{j_{1}}\cdots \delta _{i_{k}}^{j_{k}}=\frac{(N-m+k)!}{(N-m)!}\;
\delta _{[j_{k+1}\cdots j_{m}] }^{[i_{k+1}\cdots i_{m}]}\,,\quad 1\leq k\leq m\leq N\,.
$$}
\begin{eqnarray}
\mathcal{O(}1\mathcal{)} &=&\frac{\left( -1\right) ^{n}}{8\pi G(2n-2)!}
\frac{\sqrt{-h}}{\ell }\,\delta _{\left[ j_{1}\cdots j_{2n-1}\right] }^{\left[
i_{1}\cdots i_{2n-1}\right] }\delta _{i_{1}}^{j_{1}}\cdots \delta
_{i_{2n-1}}^{j_{2n-1}}\int\limits_{0}^{1}dt\left[ \left( -t^{2}\right)
^{n-1}+(-1)^{n}\right] \notag \\
&=&\frac{\left( 2n-1\right) }{8\pi G}\frac{\sqrt{-h}}{\ell }\left( 1
-\frac{1}{2n-1}\right) \notag \\
&=&\frac{\sqrt{-h}}{8\pi G}\,\frac{2n-2}{\ell }\,.
\end{eqnarray}

The term linear in the curvature comes from linear terms in the Schouten
tensor, when the rest of the indices are saturated with Kronecker deltas,
\begin{eqnarray}
\mathcal{O(R)} &=&\frac{\left( -1\right) ^{n}\ell }{8\pi G(2n-2)!}\,\sqrt{-h}\,
\delta _{\left[ j_{1}\cdots j_{2n-1}\right] }^{\left[ i_{1}\cdots i_{2n-1}%
\right] }S_{i_{1}}^{j_{1}}\delta _{i_{2}}^{j_{2}}\cdots \delta
_{i_{2n-1}}^{j_{2n-1}}\times \notag \\
&&\times \int\limits_{0}^{1}dt\left[ (-1)^{n}\left( 1-t^{2n-2}\right)
+2(-1)^{n-2}(n-1)t^{2n-4}\left( 1-t^{2}\right) \right] \notag \\
&=&\frac{\ell }{8\pi G}\,\sqrt{-h}\,S\int\limits_{0}^{1}dt\left[
1-(2n-3)t^{2n-2}+2(n-1)t^{2n-4}\right] \notag \\
&=&\frac{\ell }{8\pi G}\,\sqrt{-h}\,S\,\frac{2(n-1)}{2n-3}=\frac{\sqrt{-h}}{8\pi G}
\frac{\ell \mathcal{R}}{2(2n-3)}\,.
\end{eqnarray}
Terms linear in the Weyl tensor vanish because they involve traces of it.

One can show that $\mathcal{O}(\mathcal{R}^{2})$ terms in the expansion of
the extrinsic curvature will not affect quadratic-curvature terms in $%
\mathcal{L}(h,K,\mathcal{R})$. On the other hand, contractions of the Weyl
tensor with a single Schouten tensor will again involve traces of $\mathcal{W%
}$, such that products between $S$ and $\mathcal{W}$ are not present.
Summing up the rest of the quadratic contributions in $\mathcal{R}$, we
arrive at the expression
\begin{eqnarray}
\mathcal{O(R}^{2}\mathcal{)} &=&\frac{\left( -1\right) ^{n}\ell ^{3}}{16\pi
G(2n-3)!}\,\sqrt{-h}\,\delta _{\left[ j_{1}\cdots j_{2n-1}\right] }^{\left[
i_{1}\cdots i_{2n-1}\right] }S_{i_{1}}^{j_{1}}S_{i_{2}}^{j_{2}}\delta
_{i_{3}}^{j_{3}}\cdots \delta _{i_{2n-1}}^{j_{2n-1}}\times \notag \\
&&\times \int\limits_{0}^{1}dt\,\left( -t^{2}\right) ^{n-3}\left[
-2t^{2}\left( 1-t^{2}\right) +2(n-2)\left( 1-t^{2}\right) ^{2}+t^{4}\right] \notag \\
&=&\frac{\ell ^{3}}{16\pi G(2n-5)}\,\sqrt{-h}\,\delta _{\left[ j_{1}j_{2}\right]
}^{\left[ i_{1}i_{2}\right] }S_{i_{1}}^{j_{1}}S_{i_{2}}^{j_{2}}\,.
\end{eqnarray}

In order to obtain the standard form of the curvature-squared counterterms,
we use the identity
\begin{eqnarray}
\delta _{\left[ j_{1}j_{2}\right] }^{\left[ i_{1}i_{2}\right]
}S_{i_{1}}^{j_{1}}S_{i_{2}}^{j_{2}} &=&S^{2}-S^{ij}S_{ij}\notag \\
&=&-\frac{1}{(2n-3)^{2}}\left( \mathcal{R}_{ij}\mathcal{R}^{ij}-\frac{(2n-1)%
}{4(2n-2)}\mathcal{R}^{2}\right)\,,
\end{eqnarray}
such that
\begin{equation}
\mathcal{O(R}^{2}\mathcal{)}=\frac{\sqrt{-h}}{8\pi G}\frac{\ell ^{3}}{%
2(2n-3)^{2}(2n-5)}\left( \mathcal{R}_{ij}\mathcal{R}^{ij}-\frac{(2n-1)}{%
4(2n-2)}\mathcal{R}^{2}\right)\,.
\end{equation}

We also provide the expression for the Weyl-squared term, which is
\begin{eqnarray}
\mathcal{O(W}^{2}\mathcal{)} &=&\frac{(-1)^n\ell ^{3}}{256\pi
G(2n-3)(2n-5)!}\,\sqrt{-h}\,\delta _{\left[ j_{1}\cdots j_{2n-1}\right] }^{\left[
i_{1}\cdots i_{2n-1}\right] }\notag \\
&&\times \int\limits_{0}^{1}dt\,\left( -t^2\right) ^{n-3}\mathcal{W}%
_{i_{1}i_{2}}^{j_{1}j_{2}}\mathcal{W}_{i_{3}i_{4}}^{j_{3}j_{4}}\,\delta
_{j_{5}}^{i_{5}}\cdots \delta _{j_{2n-1}}^{i_{2n-1}} \notag \\
&=&-\frac{\ell ^{3}}{256\pi G(2n-3)(2n-5)}\,\sqrt{-h}\,\delta _{\left[
j_{1}j_{2}j_{3}j_{4}\right] }^{\left[ i_{1}i_{2}i_{3}i_{4}\right] }\mathcal{W%
}_{i_{1}i_{2}}^{j_{1}j_{2}}\mathcal{W}_{i_{3}i_{4}}^{j_{3}j_{4}}\notag  \\
&=&-\frac{\ell ^{3}}{64\pi G(2n-3)(2n-5)}\sqrt{-h}\,\mathcal{W}^{ijkl}
\mathcal{W}_{ijkl}\,.
\end{eqnarray}
This is just for the purpose of completing the computation, because the
quadratic piece in the boundary Weyl tensor has a faster asymptotic
fall-off, such that it is not considered to be part of the standard
counterterm series.

In sum, in this section we have provided a nontrivial checking that the
action defined by the addition of Kounterterms in Eq.(\ref{Irentilde}) is
equal to the renormalized AdS action,
\begin{equation}
\tilde{I}_{\text{ren}}=I_{\text{ren}}.
\end{equation}

On the other hand, the boundary term $B_{2n-1}$ appears as a boundary
correction to the Euler characteristic $\chi \left( M\right) $ in the Euler
theorem in $2n$ dimensions,
\begin{equation}
\int\limits_{M}\mathcal{E}_{2n}=(4\pi )^n\,n!\,\chi (M)
+\int\limits_{\partial M}B_{2n-1}(h,K,\mathcal{R})\,,\,
\end{equation}
where $\mathcal{E}_{2n}$ is the Euler term in that dimension. This simply
means that the GHY term plus the standard counterterm series in
$I_{\text{ren}}$ can be generated from the addition of a single topological
invariant in the bulk.

In the next section, we exploit this remarkable feature of $I_{\text{ren}}$ to work
out a general property of the on-shell value of the renormalized AdS action
in even dimensions.

%%%%%%%%%%%%%%%%%%%%%%%%%%%%%%%%%%%%%%%%%%%%%%%%%%%%%%%%%%%%%%%%%%%%%%%%%%%
\section{Renormalized action and Critical Gravity}
%%%%%%%%%%%%%%%%%%%%%%%%%%%%%%%%%%%%%%%%%%%%%%%%%%%%%%%%%%%%%%%%%%%%%%%%%%%

Let us consider the Einstein-Hilbert action with negative cosmological
constant in $D=2n$ dimensions,
\begin{equation}
I_\text{ren} =\frac{1}{16\pi G}\int d^{2n}x\sqrt{-g}\left[ R-2\Lambda +\alpha
_{2n}\,\delta _{\left[ \mu _{1}\cdots \mu _{2n}\right] }^{\left[ \nu
_{1}\cdots \nu _{2n}\right] }\,R_{\nu _{1}\nu _{2}}^{\mu _{1}\mu _{2}}\cdots
R_{\nu _{2n-1}\nu _{2n}}^{\mu _{2n-1}\mu _{2n}}\right] \,.
\end{equation}
It was shown in Ref.\cite{Mis-Ole} that the addition of the Euler term to
the even-dimensional AdS gravity action is equivalent to the Holographic
Renormalization program if the coupling constant is chosen as%
\begin{equation}
\alpha _{2n}=(-1)^{n}\frac{\ell ^{2n-2}}{2^{n}n(2n-2)!}\,.
\end{equation}
That is the reason why, from now on, we will call it renormalized action. We
can cast it in the alternative form,
\begin{eqnarray}
I_\text{ren} &=&\frac{1}{2^{n+4}\pi G(2n-2)!}\int d^{2n}x\sqrt{-g}\,
\delta_{\left[ \mu _{1}\cdots \mu _{2n}\right] }^{\left[ \nu _{1}\cdots \nu _{2n}
\right] }\,\left[ \rule{0pt}{17pt} R_{\nu _{1}\nu _{2}}^{\mu _{1}\mu _{2}}\,\delta _{\left[ \nu
_{3}\nu _{4}\right] }^{\left[ \mu _{3}\mu _{4}\right] }\cdots \delta _{\left[
\nu _{2n-1}\nu _{2n}\right] }^{\left[ \mu _{2n-1}\mu _{2n}\right] }
\right.  \notag \\
&&+\frac{n-1}{n\ell ^{2}}\,\delta _{\left[ \nu _{1}\nu _{2}\right] }^{\left[ \mu _{1}\mu
_{2}\right] }\cdots \delta _{\left[ \nu _{2n-1}\nu _{2n}\right] }^{\left[
\mu _{2n-1}\mu _{2n}\right] }
+\left. \frac{(-1)^n}{n}\,\ell^{2n-2}\,R_{\nu _{1}\nu _{2}}^{\mu _{1}\mu
_{2}}\cdots R_{\nu _{2n-1}\nu _{2n}}^{\mu _{2n-1}\mu _{2n}}\rule{0pt}{15pt}
\right].
\label{IrenEven}
\end{eqnarray}
Now, we use the fact that, on-shell, the Weyl tensor is
\begin{equation}
W_{\mu \nu }^{\alpha \beta }=R_{\mu \nu }^{\alpha \beta }+\frac{1}{\ell ^{2}}
\,\delta _{\left[ \mu \nu \right] }^{\left[ \alpha \beta \right] }\,,
\label{AdScurvature}
\end{equation}
such that we replace this relation in $I_{\text{ren}}$ and we get
\begin{eqnarray}
I_{\text{ren}} &=&\frac{1}{2^{n+4}\pi G(2n-2)!}\int d^{2n}x\,\sqrt{-g}\,\delta
_{\left[ \mu _{1}\cdots \mu _{2n}\right] }^{\left[ \nu _{1}\cdots \nu _{2n}
\right] }\times \notag \\
&&
\times\left[ W_{\nu _{1}\nu _{2}}^{\mu _{1}\mu _{2}}\,\delta _{\left[ \nu
_{3}\nu _{4}\right] }^{\left[ \mu _{3}\mu _{4}\right] }\cdots \delta _{\left[
\nu _{2n-1}\nu _{2n}\right] }^{\left[ \mu _{2n-1}\mu _{2n}\right] }
 -\frac{1}{n\ell ^{2}}\,\delta _{\left[ \nu _{1}\nu _{2}\right] }^{\left[ \mu _{1}\mu
_{2}\right] }\cdots \delta _{\left[ \nu _{2n-1}\nu _{2n}\right] }^{\left[
\mu _{2n-1}\mu _{2n}\right]} + \right.
\notag \\
&&+\left. \frac{(-1)^{n}}{n}\,\ell ^{2n-2}\left( W_{\nu _{1}\nu _{2}}^{\mu
_{1}\mu _{2}}-\frac{1}{\ell ^{2}}
\delta _{\left[ \nu _{1}\nu _{2}\right] }^{\left[ \mu _{1}\mu _{2}\right]}\right)
\cdots \left( W_{\nu _{2n-1}\nu_{2n}}^{\mu _{2n-1}\mu _{2n}}
-\frac{1}{\ell ^{2}}\delta _{\left[ \nu_{2n-1}\nu _{2n}\right] }^{\left[ \mu _{2n-1}\mu _{2n}\right] }\right)
\right] \,.  \label{IrenW}
\end{eqnarray}
Expanding the binomial in the last line, we obtain
\begin{eqnarray}
&&
\frac{(-1)^{n}}{n}\,\ell ^{2n-2}\,\delta _{\left[ \mu _{1}\cdots \mu _{2n}
\right] }^{\left[ \nu _{1}\cdots \nu _{2n}\right] }\left( W_{\nu _{1}\nu
_{2}}^{\mu _{1}\mu _{2}}-\frac{1}{\ell ^{2}}\,\delta _{\left[ \nu _{1}\nu _{2}%
\right] }^{\left[ \mu _{1}\mu _{2}\right] }\right) \cdots \left( W_{\nu
_{2n-1}\nu _{2n}}^{\mu _{2n-1}\mu _{2n}}-\frac{1}{\ell ^{2}}\,\delta _{\left[
\nu _{2n-1}\nu _{2n}\right] }^{\left[ \mu _{2n-1}\mu _{2n}\right] }\right) \notag\\
&& =
\delta _{\left[ \mu _{1}\cdots \mu _{2n}\right] }^{\left[ \nu _{1}\cdots
\nu _{2n}\right] }\left( \frac{1}{n\ell ^{2}}\,\delta _{\left[ \nu _{1}\nu
_{2}\right] }^{\left[ \mu _{1}\mu _{2}\right] }\cdots \delta _{\left[ \nu
_{2n-1}\nu _{2n}\right] }^{\left[ \mu _{2n-1}\mu _{2n}\right] }-W_{\nu
_{1}\nu _{2}}^{\mu _{1}\mu _{2}}\delta _{\left[ \nu _{3}\nu _{4}\right] }^{%
\left[ \mu _{3}\mu _{4}\right] }\cdots \delta _{\left[ \nu _{2n-1}\nu _{2n}%
\right] }^{\left[ \mu _{2n-1}\mu _{2n}\right] } + \right. \notag\\
&&
+\left. \frac{\ell ^{2}}{2}
\,(n-1)W_{\nu _{1}\nu _{2}}^{\mu _{1}\mu _{2}}W_{\nu _{3}\nu _{4}}^{\mu
_{3}\mu _{4}}\cdots \delta _{\left[ \nu _{2n-1}\nu _{2n}\right] }^{\left[
\mu _{2n-1}\mu _{2n}\right] } \right)+ \mathcal{O}(W^3)\,.
\end{eqnarray}
The first term in the above expansion cancels the second term in the first
line of Eq.(\ref{IrenW}). All terms linear in the Weyl tensor vanish because
they involve traces of it. As a consequence, the first non-vanishing
contribution in the renormalized action is quadratic in $W$,
\begin{eqnarray}
I_{\text{ren}} &=&\frac{\ell ^{2}}{2^{n+6}\pi G(2n-3)!}\int d^{2n}x\sqrt{-g}\,
\delta _{\left[ \mu _{1}\cdots \mu _{2n}\right] }^{\left[ \nu _{1}\cdots \nu
_{2n}\right] } W_{\nu _{1}\nu _{2}}^{\mu _{1}\mu
_{2}}W_{\nu _{3}\nu _{4}}^{\mu _{3}\mu _{4}}\cdots \delta _{\left[ \nu
_{2n-1}\nu _{2n}\right] }^{\left[ \mu _{2n-1}\mu _{2n}\right] }+ \mathcal{O}(W^3).
\end{eqnarray}
We can also write it as
\begin{equation}
I_{\text{ren}} =\frac{\gamma_{_\text{CG}}}{16\pi G} \int d^{2n}x\sqrt{-g}\,
W^{\alpha \beta \mu \nu }\,W_{\alpha \beta \mu \nu }+ \mathcal{O}(W^3)\,,
\label{mainresult}
\end{equation}
because the coupling,
\begin{equation}
\gamma_{_\text{CG}} =\frac{\ell ^{2}}{4(2n-3)}=-\frac{(2n-1)(2n-2)}{8\Lambda (2n-3)}\,,
\end{equation}
is the same one that appears in the Critical Gravity action (\ref{CGA}).

%%%%%%%%%%%%%%%%%%%%%%%%%%%%%%%%%%%%%%%%%%%%%%%%%%%%%%%%%%%%%%%%%%%%%%%%%%%
\section{Conclusions}
%%%%%%%%%%%%%%%%%%%%%%%%%%%%%%%%%%%%%%%%%%%%%%%%%%%%%%%%%%%%%%%%%%%%%%%%%%%

We have shown that, in even spacetime dimensions, the renormalized AdS
action is on-shell equivalent to a polynomial of the Weyl tensor, whose
first nonvanishing contribution is $Weyl^2$. The coupling of this term is
the same as the one that appears in Critical Gravity, where $Weyl^2$
term is added on top of the Einstein-Hilbert Lagrangian.

We stress that this equivalence is at the level of the action evaluated for
Einstein spacetimes and, by no means, we imply a dynamic equivalence between
the corresponding theories.

We also emphasize that the fact $I_\text{ren}=\frac{\gamma _{_\text{CG}}}{16\pi
G}\,Weyl^2+\cdots$ is a consequence of a \emph{topological} regularization
of AAdS gravity. This can only be seen once one shows that the addition of
topological invariants and Holographic Renormalization program in even-dimensional AdS
gravity provide the same result.

Because of this argument, it is difficult to think of a similar
result for odd-dimensional case.

We can go back to four-dimensional example worked out by Lu and Pope in
Ref.\cite{Lu:2011zk}, in order to see what the above
claim implies in that case. In 4D
Critical Gravity, the action has the form
\begin{equation}
I_\text{CG}=\frac{1}{16\pi G}\int d^{4}x\,\sqrt{-g}\left[ \left( R+%
\frac{1}{\alpha }\right) +\alpha R^{2}-3\alpha R_{\mu \nu }R^{\mu \nu }%
\right] .
\end{equation}
When the cosmological term adopts the standard value of Einstein-AdS gravity
($\alpha =\ell ^{2}/6$), the action can be rewritten as
\begin{equation}
I_\text{CG}=\frac{1}{16\pi G}\int d^{4}x\,\sqrt{-g}\left[ \left( R+%
\frac{6}{\ell ^{2}}\right) -\frac{\ell ^{2}}{2}\left( R_{\mu \nu }R^{\mu \nu
}-\frac{1}{3}R^{2}\right) \right] .
\end{equation}

Quadratic terms in the curvature are given just as the difference between $%
Weyl^{2}$ and Gauss-Bonnet terms
\begin{equation}
I_\text{CG}=\frac{1}{16\pi G}\int d^{4}x\,\sqrt{-g}\left[ \left( R+%
\frac{6}{\ell ^{2}}\right) -\frac{\ell ^{2}}{4}\left( W^{2}-GB\right) \right].
\end{equation}
 The particular coupling of Gauss-Bonnet term, as originally pointed out in Ref.\cite{Mis-Ole},
 leads to the renormalized AdS action given by
Eq.(\ref{IrenEven}), such that the total action for Critical Gravity is%
\begin{equation}
I_\text{CG}=I_\text{ren}-\frac{\ell ^{2}}{64\pi G}\int d^{4}x\,\sqrt{-g}\,
W^{2}\,.
\end{equation}

Notice that this automatically implies that $I_\text{CG}=0$ for Einstein spaces,
what seems to indicate that the critical point defines a new vacuum state of
the theory.

In higher even dimensions, going beyond quadratic terms in the Weyl tensor in the expansion of the
renormalized action (\ref{IrenW}), we get
\begin{eqnarray}
I_{\text{ren}}\  &=&\frac{\gamma _{_{\text{CG}}}}{16\pi G}\int d^{2n}x\sqrt{%
-g}\,W^{\alpha \beta \mu \nu }W_{\alpha \beta \mu \nu } - \notag \\
&&-\frac{\ell ^{4}\left( n-2\right) }{2^{n+6}3\pi G(2n-3)!}\int d^{2n}x\sqrt{%
-g}\,\delta _{\lbrack \mu _{1}\cdots \mu _{2n}]}^{[\nu _{1}\cdots \nu _{2n}]}%
\left[ \rule{0pt}{17pt}W_{\nu _{1}\nu _{2}}^{\mu _{1}\mu _{2}}W_{\nu _{3}\nu
_{4}}^{\mu _{3}\mu _{4}}W_{\nu _{5}\nu _{6}}^{\mu _{5}\mu _{6}}\,\delta
_{\lbrack \nu _{7}\nu _{8}]}^{[\mu _{7}\mu _{8}]}\cdots \delta _{\left[ \nu
_{2n-1}\nu _{2n}\right] }^{\left[ \mu _{2n-1}\mu _{2n}\right] } + \right.
\notag \\
&&\,+6\left( n-3\right) !\left. \sum_{p\geq 4}^{n}\frac{\left( -1\right)
^{p+1}\ell ^{2p-6}}{p!\left( n-p\right) !}\,W_{\nu _{1}\nu _{2}}^{\mu
_{1}\mu _{2}}\cdots W_{\nu _{2p-1}\nu _{2p}}^{\mu _{2p-1}\mu _{2p}}\delta _{%
\left[ \nu _{2p+1}\nu _{2p+2}\right] }^{\left[ \mu _{2p+1}\mu _{2p+2}\right]
}\cdots \delta _{\left[ \nu _{2n-1}\nu _{2n}\right] }^{\left[ \mu _{2n-1}\mu
_{2n}\right] }\right] \,.  \label{Irengen}
\end{eqnarray}
Note that this action is not Weyl invariant even though it is expressed
on-shell in terms of the Weyl tensor. Namely, under the Weyl transformations
$g_{\mu \nu }\rightarrow \Omega ^{2}(x)\,g_{\mu \nu }$, the tensor
$W^\mu _{\;\;\nu\alpha \beta }$ is invariant, but $W_{\alpha \beta }^{\mu \nu }$
changes as $W_{\alpha \beta }^{\mu \nu }\rightarrow \Omega ^{-2}W_{\alpha
\beta }^{\mu \nu }$ . Taking into consideration that the volume element also
transforms as $d^{2n}x\sqrt{-g}\rightarrow d^{2n}x\sqrt{-g}\,\Omega ^{2n}$,
we find that the $p$-th term of the polynomial in $W$ transforms with the
weight $n-p$,
\begin{equation}
\left( \sqrt{-g}\,\delta _{\left[ \mu _{1}\cdots \mu _{2p}\right] }^{\left[
\nu _{1}\cdots \nu _{2p}\right] }W_{\nu _{1}\nu _{2}}^{\mu _{1}\mu
_{2}}\cdots W_{\nu _{2p-1}\nu _{2p}}^{\mu _{2p-1}\mu _{2p}}\right)
\rightarrow \Omega ^{2n-2p}\left( \sqrt{-g}\,\delta _{\left[ \mu _{1}\cdots
\mu _{2p}\right] }^{\left[ \nu _{1}\cdots \nu _{2p}\right] }W_{\nu _{1}\nu
_{2}}^{\mu _{1}\mu _{2}}\cdots W_{\nu _{2p-1}\nu _{2p}}^{\mu _{2p-1}\mu
_{2p}}\right).
\end{equation}
In particular, in $D=4$, there is only one term with $n=$ $p=2$, thus this
theory is Weyl invariant.

The expression (\ref{Irengen}) can be rearranged as
\begin{eqnarray}
I_{\text{ren}}\  &=&\frac{\gamma _{_{\text{CG}}}}{16\pi G}\int d^{2n}x\sqrt{%
-g}\left[ \rule{0pt}{17pt}\,W^{\alpha \beta \mu \nu }W_{\alpha \beta \mu \nu
}\right. +a\,\left( W_{\alpha \beta }^{\mu \nu }W_{\lambda \rho }^{\alpha
\beta }W_{\mu \nu }^{\lambda \rho }-4W_{\rho \beta }^{\mu \nu }W_{\nu
\lambda }^{\alpha \beta }W_{\mu \alpha }^{\lambda \rho }\right) +  \notag \\
&&\,+\left. \sum_{p\geq 4}^{n}b_{p}\,\delta _{\left[ \mu _{1}\cdots \mu _{2p}%
\right] }^{\left[ \nu _{1}\cdots \nu _{2p}\right] }\,W_{\nu _{1}\nu
_{2}}^{\mu _{1}\mu _{2}}\cdots W_{\nu _{2p-1}\nu _{2p}}^{\mu _{2p-1}\mu
_{2p}}\right] \,,
\end{eqnarray}
using the identity for the cubic term in $W$,
\begin{equation}
\delta _{\left[ \mu _{1}\cdots \mu _{6}\right] }^{\left[ \nu _{1}\cdots \nu
_{6}\right] }\,W_{\nu _{1}\nu _{2}}^{\mu _{1}\mu _{2}}W_{\nu _{3}\nu
_{4}}^{\mu _{3}\mu _{4}}W_{\nu _{5}\nu _{6}}^{\mu _{5}\mu _{6}}=2^{4}\left(
W_{\alpha \beta }^{\mu \nu }W_{\lambda \rho }^{\alpha \beta }W_{\mu \nu
}^{\lambda \rho }-4W_{\alpha \lambda }^{\mu \nu }W_{\mu \rho }^{\alpha \beta
}W_{\nu \beta }^{\lambda \rho }\right) \,.
\end{equation}
The corresponding couplings of $Weyl^3$ and all higher-order terms are
\begin{eqnarray}
a &=&2^{4}b_{3}=-\frac{\ell ^{2}}{3\left( 2n-5\right) }\,,  \notag \\
b_{p} &=&\left( -1\right) ^{p}\ell ^{2p-4}\frac{\left( n-2\right) !\left(
2n-2p\right) !}{2^{p-1}\left( 2n-4\right) !\,p!\left( n-p\right) !}\,,\quad
p\geq 3\,.
\end{eqnarray}
For the purpose of comparison with Critical Gravity with cubic-curvature
contributions developed in Ref.\cite{Sisman:2011gz}, we use the definition
of the Weyl tensor
\begin{equation}
W_{\alpha \beta }^{\mu \nu }=R_{\alpha \beta }^{\mu \nu }-\left( \delta
_{\alpha }^{\mu }S_{\beta }^{\nu }-\delta _{\alpha }^{\nu }S_{\beta }^{\mu
}-\delta _{\beta }^{\mu }S_{\alpha }^{\nu }+\delta _{\beta }^{\nu }S_{\alpha
}^{\mu }\right) \,,
\end{equation}
in terms of the spacetime Schouten tensor
\begin{equation}
S_{\mu }^{\nu }=\frac{1}{D-2}\left( R_{\mu }^{\nu }-\frac{1}{2\left(
D-1\right) }\,\delta _{\mu }^{\nu }\,R\right) \,.
\end{equation}
In doing so, we obtain
\begin{eqnarray}
&&\frac{1}{16}\,\delta _{\left[ \mu _{1}\cdots \mu _{6}\right] }^{\left[ \nu
_{1}\cdots \nu _{6}\right] }\,W_{\nu _{1}\nu _{2}}^{\mu _{1}\mu _{2}}W_{\nu
_{3}\nu _{4}}^{\mu _{3}\mu _{4}}W_{\nu _{5}\nu _{6}}^{\mu _{5}\mu _{6}}
=R_{\alpha \beta }^{\mu \nu }R_{\lambda \rho
}^{\alpha \beta }R_{\mu \nu }^{\lambda \rho }-4R_{\alpha \beta }^{\mu \nu }R_{\mu \rho }^{\alpha \lambda
}R_{\nu \lambda }^{\beta \rho }-\frac{36}{D-2}\,R_{\alpha \beta }^{\mu \nu }R_{\mu \nu }^{\alpha \lambda
}R_{\lambda }^{\beta } \notag \\
&& \frac{18}{\left( D-1\right) \left( D-2\right) }\,RR_{\alpha \beta }^{\mu \nu }R_{\mu \nu
}^{\alpha \beta }+\frac{12\left( D+4\right) }{\left( D-2\right)^{2}}
\,R_{\alpha \beta }^{\mu \nu }R_{\mu }^{\alpha }R_{\nu }^{\beta }
+\frac{8\left( 7D-8\right) }{\left( D-2\right) ^{3}}\,R_{\nu }^{\mu
}R_{\lambda }^{\nu }R_{\mu }^{\lambda }  \notag \\
&&-\frac{12\left( D^{2}+9D-16\right) }{\left( D-1\right) \left( D-2\right)
^{3}}\left( RR_{\nu }^{\mu }R_{\mu }^{\nu }-\frac{1}{3\left( D-1\right) }
\,R^{3}\right) .
\end{eqnarray}
Found cubic gravity belongs to a class of cubic critical gravities discussed in Ref.\cite{Sisman:2011gz}. There, all gravitational theories with up to cubic curvature terms were classified based on the requirement of unitarity around (A)dS vacuum. However, conditions of criticality (removal of the massive spin-0 mode and also that the spin-2 mode be massless) fix only two of eight cubic coupling constants in terms of the others.

Regarding the result given by Eq.(\ref{mainresult}), at this moment, we
cannot further understand the implications of this remarkable feature of the
renormalized AdS action.

However, holographic renormalization method applied to the theory around the critical point
in the action of Critical Gravity may give some insight on this problem.
AAdS spaces which are solutions of the Einstein equations are described by
the Fefferman-Graham metric \cite{Fefferman-Graham}. Higher-derivative terms in the field
equations imply the existence of new holographic sources at the boundary,
which should appear at a given order in the asymptotic expansion of the
metric. Significant progress towards a holographic description of Critical Gravity
has been made in four dimensions in Ref.\cite{Johansson:2012fs}, where
logarithmic modes play an important role.
The main result presented here, Eq.(\ref{mainresult}), seems to
indicate the exact cancelation of Einstein modes in the metric of a
spacetime which is a solution to Critical Gravity. Therefore, the residual
dynamics should be given just in terms of the new sources of the full theory.

%%%%%%%%%%%%%%%%%%%%%%%%%%%%%%%%%%%%%%%%%%%%%%%%%%%%%%%%%%%%%%%%%%%%%%%%%%%
\acknowledgments

We thank T.C. Sisman and G. Giribet for useful discussions on critical
gravities. M.T. would like to thank the Pontificia Universidad Cat\'olica de
Valpara\'\i so and the Universidad Andr\'es Bello for the kind hospitality during the
initial stages of this work. This work was supported by the Chilean FONDECYT
Grants No.1110102, No.1131075 and No.3120143. O.M. also thanks DII-PUCV
for support through the project No.123.711/2011. The work of R.O. is
financed in part by the UNAB grant DI-551-14/R. The Centro de Estudios
Cientificos (CECs) is funded by the Chilean Government through the Centers
of Excellence Base Financing Program of CONICYT.

%%%%%%%%%%%%%%%%%%%%%%%%%%%%%%%%%%%%%%%%%%%%%%%%%%%%%%%%%%%%%%%%%%%%%%%%%%%

\end{document}